\documentclass[conference]{IEEEtran}
\IEEEoverridecommandlockouts

\newcommand{\subparagraph}{}
\usepackage{algpseudocode}
\usepackage{amsmath}
\usepackage{amsmath,amssymb,amsfonts}
\usepackage{array}
\usepackage{balance}
\usepackage{blindtext}
\usepackage{cite}
\usepackage{comment}
\usepackage[english]{babel}
\usepackage{enumitem} %
\usepackage{floatrow}
\usepackage{flushend}
\usepackage[font=footnotesize]{caption}
\usepackage{graphics}
\usepackage{graphicx}
\usepackage{lipsum, color}
\usepackage{mathtools}
\usepackage{multicol}
\usepackage{multirow}
\usepackage{siunitx}
\usepackage{soul}
\usepackage{subcaption}
\usepackage{subfig}
\usepackage[T1]{fontenc}
\usepackage{tabularx}
\usepackage{textcomp}
\usepackage{tikz}
\usepackage{titlesec}
\usepackage{url}

\IEEEoverridecommandlockouts

\usepackage{siunitx}

\def\BibTeX{{\rm B\kern-.05em{\sc i\kern-.025em b}\kern-.08em
    T\kern-.1667em\lower.7ex\hbox{E}\kern-.125emX}}

\begin{document}
\renewcommand{\figurename}{Fig.}

\title{Smart Pixels: In-pixel AI for on-sensor data filtering\vspace{-10pt}}

\author{
\small
Benjamin Parpillon$^{1,2}$, 
Chinar Syal$^1$, 
Jieun Yoo$^2$,
Jennet Dickinson$^1$,
Morris Swartz$^3$,
Giuseppe Di Guglielmo$^{1,4}$,\\
Alice Bean$^5$,
Douglas Berry$^1$, 
Manuel Blanco Valentin$^4$,
Karri DiPetrillo$^6$,
Anthony Badea$^6$,
Lindsey Gray$^2$,\\
Petar Maksimovic$^3$,
Corrinne Mills$^2$,
Mark S. Neubauer$^8$,
Gauri Pradhan$^1$,
Nhan Tran$^{1,4}$, 
Dahai Wen$^{3}$,
Farah Fahim$^{1}$\\
{\scriptsize $^{1}$Fermi National Accelerator Laboratory (FNAL), $^{2}$University of Illinois Chicago, $^3$ Johns Hopkins University, $^4$ Northwestern University,}\\{\scriptsize $^5$ University of Kansas, $^6$ The University of Chicago, $^8$ University of Illinois Urbana-Champaign, Champaign}\vspace{-12pt}}

\maketitle

\begin{abstract}
We present a smart pixel prototype readout integrated circuit (ROIC) designed in CMOS 28 nm bulk process, with in-pixel implementation of an artificial intelligence (AI) / machine learning (ML) based data filtering algorithm designed as proof-of-principle for a Phase III upgrade at the Large Hadron Collider (LHC) pixel detector. 
The first version of the ROIC consists of two matrices of 256 smart pixels, each 25$\times$25 µm\textsuperscript{2} in size. Each pixel consists of a charge-sensitive preamplifier with leakage current compensation and three auto-zero comparators for a 2-bit flash-type ADC. The frontend is capable of synchronously digitizing the sensor charge within 25 ns. Measurement results show an equivalent noise charge (ENC) of $\sim$30e\textsuperscript{-} and a total dispersion of $\sim$100e\textsuperscript{-}
The second version of the ROIC uses a fully connected two-layer neural network (NN) to process information from a cluster of 256 pixels to determine if the pattern corresponds to highly desirable high-momentum particle tracks for selection and readout. The digital NN is embedded in-between analog signal processing regions of the 256 pixels without increasing the pixel size and is implemented as fully combinatorial digital logic to minimize power consumption and eliminate clock distribution, and is active only in the presence of an input signal. The total power consumption of the neural network is $\sim$ 300 $\mu$W.
The NN performs momentum classification based on the generated cluster patterns and even with a modest momentum threshold, it is capable of 54.4\%  – 75.4\%  total data rejection, opening the possibility of using the pixel information at 40MHz for the trigger. The total power consumption of analog and digital functions per pixel is $\sim$ 6 $\mu$W per pixel, which corresponds to $\sim$ 1 W/cm\textsuperscript{2} staying within the experimental constraints. 

\end{abstract}

\section{Introduction}
Detectors at forthcoming high-energy colliders will confront substantial technical hurdles. Managing the unprecedented volume of particles anticipated in each event demands highly detailed silicon pixel detectors with billions of readout channels. With event frequencies reaching 40 MHz, these detectors will produce immense amounts of data every second. To facilitate discoveries while adhering to stringent bandwidth and latency constraints, future trackers must possess the ability to swiftly, efficiently, and robustly reduce data at the source while operating in high radiation environments. This prototype development targets Phase III pixel detector upgrades projected for 2034, as well as other future colliders.

\section {Analog Frontend for Hit Detection and Digitization with Test results}
Reducing the pixel sizes by a factor of four while maintaining the same sensor thickness leads to charge sharing which can be positively exploited to improve the effective position resolution \cite{MonchBergamaschi}. However, this requires reducing the minimum detection threshold of the frontend and overcoming pileup errors.

\begin{figure}[t!]
  \centering
  \includegraphics[width=0.75\linewidth]{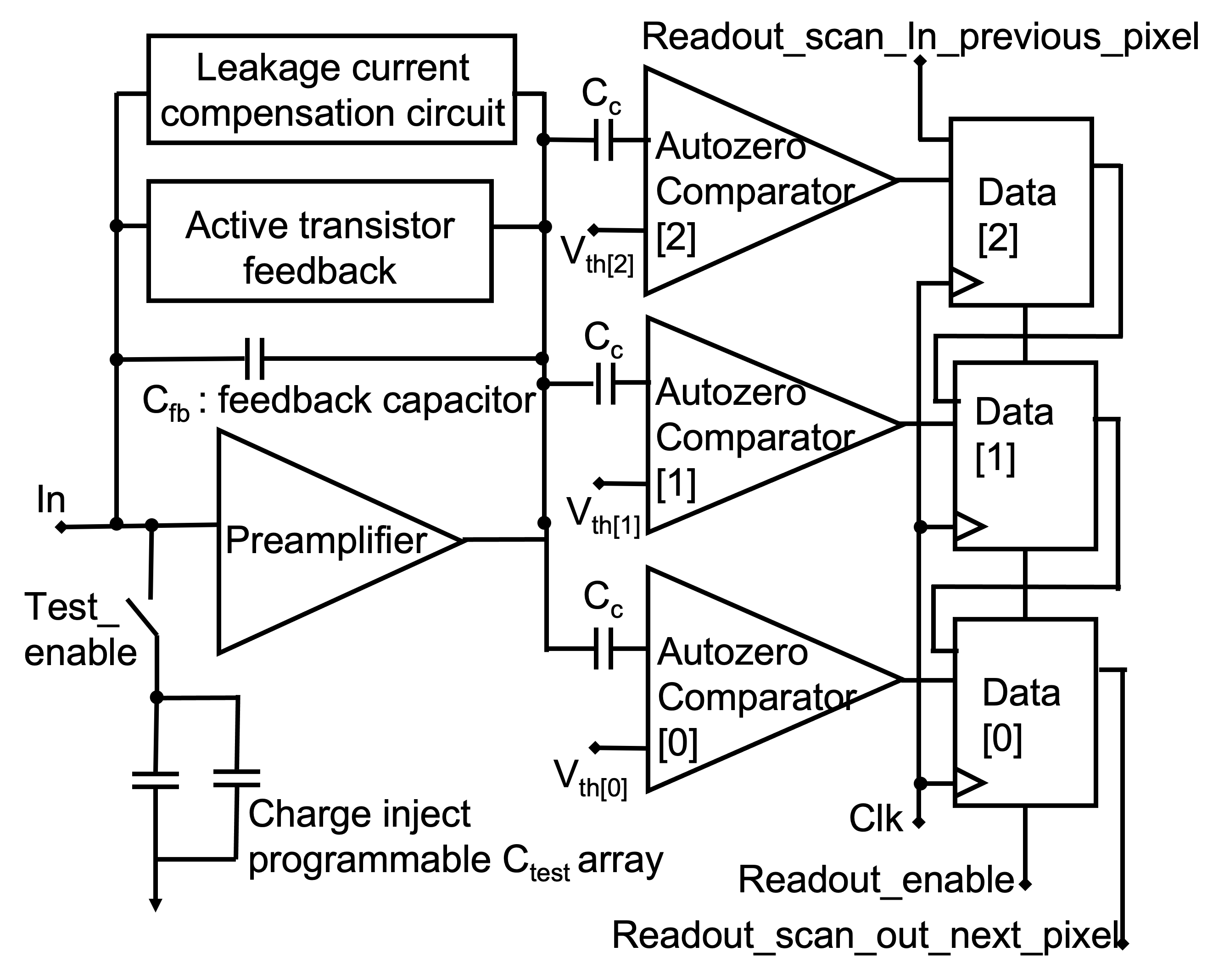}
  \caption{The pixel analog frontend for hit registration in a particle detector containing a preamplifier with leakage current compensation and AC coupled synchronous comparators.}
  \label{fig:chip_block_diagram}
\end{figure}

\subsection{Architecture}
The pixel architecture for in-pixel detection and digitization of collision events is depicted in Fig. \ref{fig:chip_block_diagram}.

The charge collected at the sensor's electrode is integrated, amplified, and converted to voltage using a charge-sensitive preamplifier. An AC-coupled 2-bit flash-type ADC digitizes the signal. Due to the thermometric nature of the flash ADC in our design, analog-to-digital conversion begins as soon as the integrated charge output is above the first threshold, and continues until the signal reaches its maximum value or the time for conversion runs out (since all comparators are reset at the end of the bunch crossing cycle). 
The first ROIC prototype was designed and fabricated in 2023. The topology, implementation and simulation results are discussed in detail in \cite{ISCAS2023}. 

\begin{figure}
  \centering
  \includegraphics[width=1\linewidth]{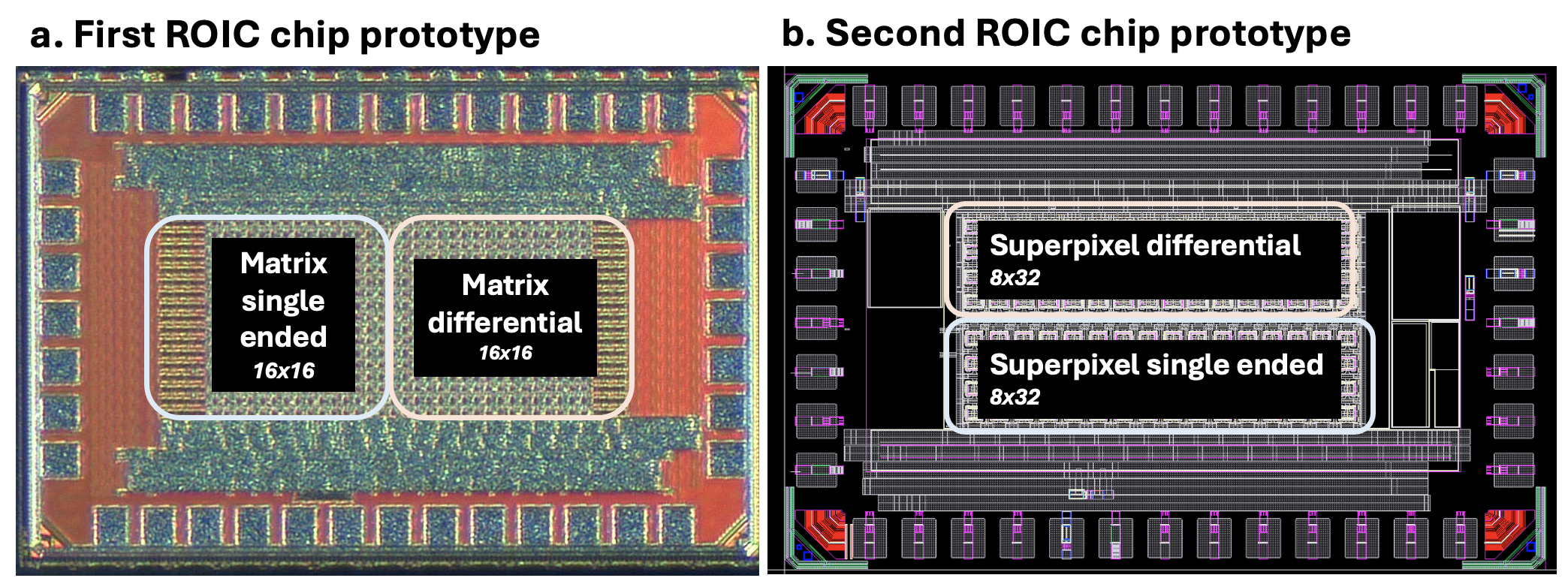}
  \caption{a. Picture of the  first 1.5mm$^2$ ROIC chip. The chip consists of two 16 × 16-pixel matrix variants with just the analog frontend. b. Layout of the second prototype chip. The chip consists of two 8 × 32 - pixel matrix variants with the analog frontend and the digital NN}
  \label{fig:CMSPIXv1_pic}
\end{figure}

\subsection{Test Results Of First ROIC Prototype}
\begin{figure}
  \centering
  \includegraphics[width=0.95\linewidth]{
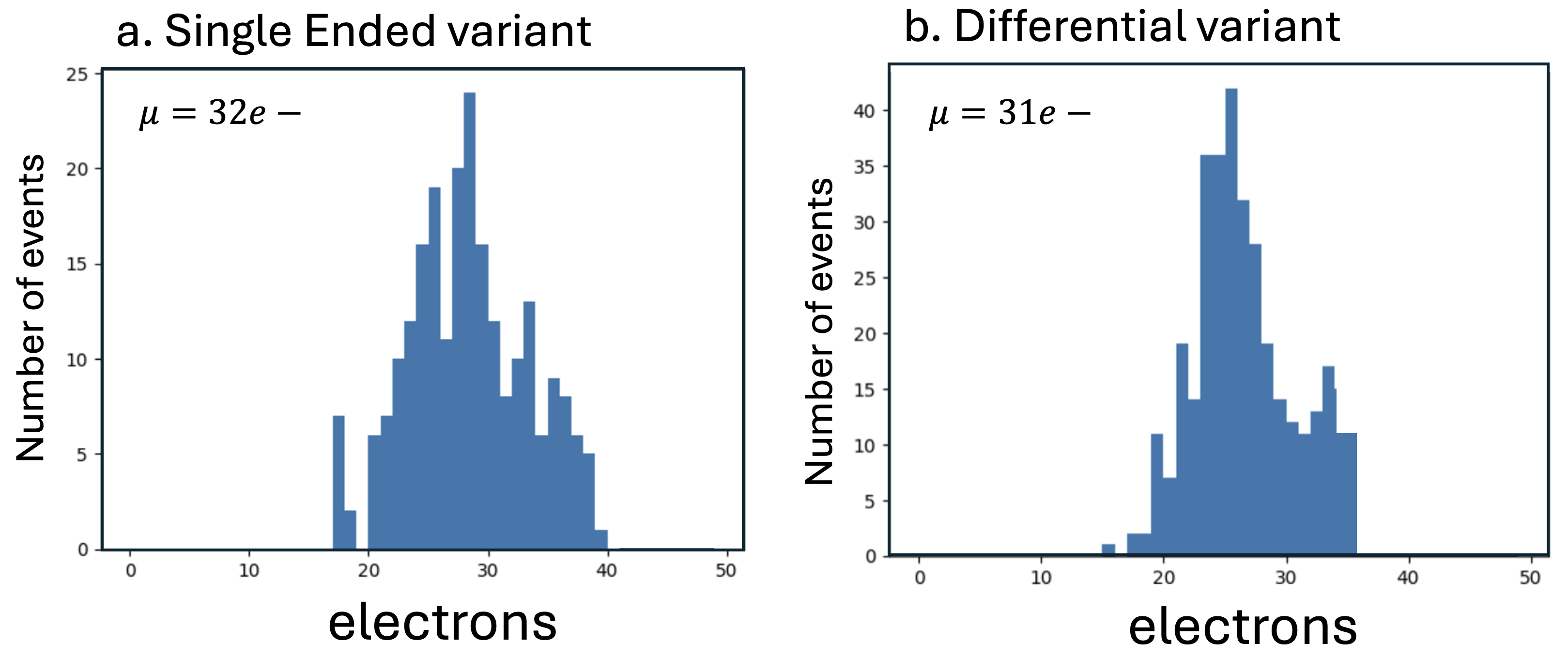}
  \caption{Total pixel ENC for both variant with 400e- equivalent threshold voltages and no sensor capacitance connected to the ROIC. a. ENC histogram of the single ended variant. b. ENC histogram of the differential variant}
  \label{fig:noise}
\end{figure}

\begin{figure}
  \centering
  \includegraphics[width=0.95\linewidth]{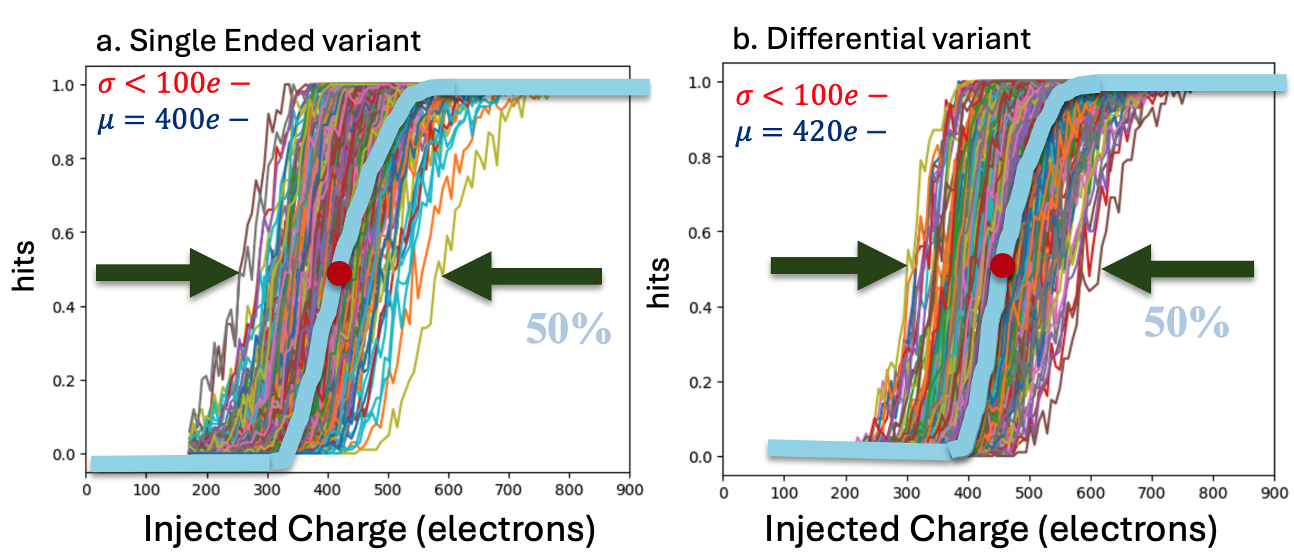}
  \caption{S-curves of total dispersion for all Hit Comparators in the Array for both variants with 400e- equivalent threshold voltages and no sensor capacitance connected to the ROIC}
  \label{fig:Scurve}
\end{figure}

Fig. \ref{fig:CMSPIXv1_pic}a shows the first ASIC chip, \SI{1.5}{\milli\meter^2} in size. It is composed of a 512 pixels matrix divided into two halves with a single-ended and differential frontend variants that were implemented. The ROIC was characterized without connecting any sensor and all the comparators in the pixel were set at 400e- equivalent threshold voltages. The ASIC incorporates programmable front-end charge injection circuitry to generate pixel cluster charges during characterization (bottom left in Fig. \ref{fig:chip_block_diagram}). The programming didn't work as reliably as expected which added a stochastic error to our results. The total equivalent noise charge (ENC) of the preamplifier and the comparator is presented in Fig. \ref{fig:noise}. The histograms were extracted by scanning the entire input dynamic range in steps of 2 e- and injecting charges 200 times for each step at every pixel. After removing the incorrectly programmed pixels, we find the ENC to be 32e- for both variants.
The total dispersion for each matrix variant is presented in Fig. \ref{fig:Scurve}. For both variants, the effective threshold dispersion of the hit comparator across the entire ASIC is 100 e\textsuperscript{-}. This result includes gain dispersion in the preamplifier and threshold dispersion of the comparators. 

\section{On-chip Neural Network Implementation}

A neural network classifier was designed and implemented on the second ROIC prototype (Fig. \ref{fig:CMSPIXv1_pic}b) to identify clusters associated with high $p_T$ charged particles. The high $p_T$ signal class contains tracks with $p_T>200$ MeV. Oppositely charged particle tracks curve in opposite directions in the magnetic field, resulting in clusters of very different shape, therefore two background classes are defined, corresponding to positively and negatively charged particles with $p_T<200$\,MeV.  The choice of $p_T$ threshold in the training class definition affects the physics performance and could be adjusted depending on the physics goals. The details about the training datasets and the algorithm development can be found in this reference \cite{yoo2023smart}.

\begin{figure}[htbp]
  \centering
  \includegraphics[width=1.0\linewidth]{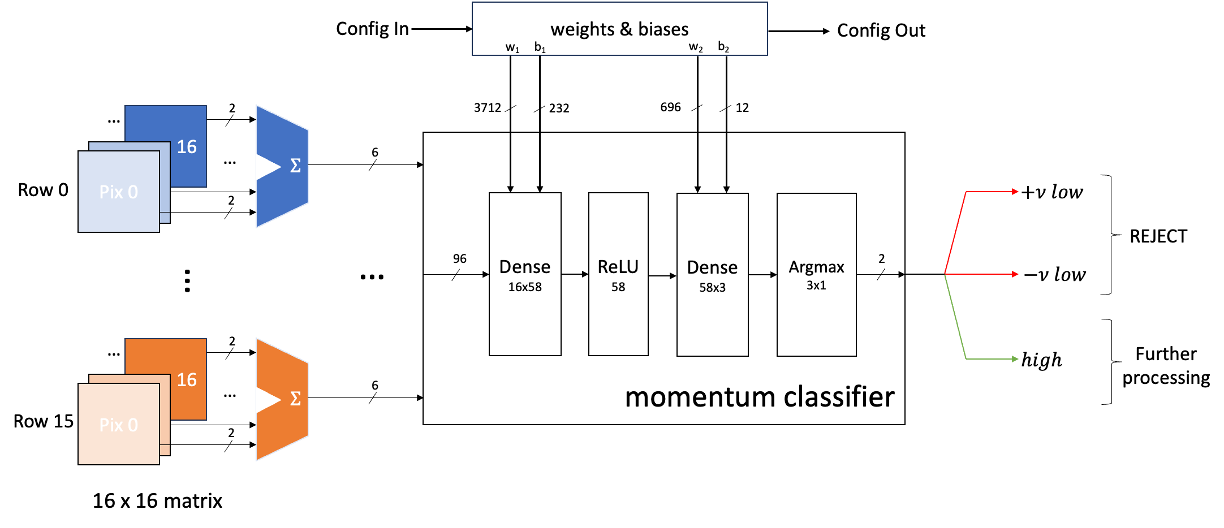}
   \caption{Data flow through the digital implementation of the algorithm from the summed ADC bits (on the left) through the neural network and the final classification layer. At the top of the diagram we illustrate the reconfigurability of the weights and biases in the algorithm stored in memory.}
  \label{fig:flow}
\end{figure}

We adopted \texttt{hls4ml} to translate the neural network classifier into optimal hardware implementations~\cite{fahim2021hls4ml}.
\texttt{hls4ml} is an open-source Python framework that facilitates the co-design of machine learning algorithms for hardware deployment, supporting models from quantized models from QKeras and other formats~\cite{coelho2021automatic}. hls4ml allowed us to fine-tune the numerical precision and the hardware parallelism to optimize area, performance, and power consumption according to our system constraints.
The conversion process started with the quantized model of the classifier, which \texttt{hls4ml} translated into HLS-ready C++ code for Siemens Catapult HLS~\cite{catapult-hls} that generates a hardware description at the register-transfer level (RTL) suitable for the ASIC flow. We chose to fully parallelize the hardware logic to minimize the latency of the neural network, integrating the HLS-generated RTL design with system registers and data movers for efficient operation shown in Fig.~\ref{fig:flow}.

\section{Conclusions and Future Works}

We have presented the first prototype test results and showed the analog frontend based on a synchronous ADC architecture is capable of processing and digitizing input signals within a 25 ns bunch crossing period. The compact area of the analog part of the pixel (<210µm\textsuperscript{2}) results in $\times$4 improved granularity while offering a $\times$2.5 power reduction ($\sim$4 µW/pixel) and an $\times$2.5 improvement in the minimum threshold detection ($\approx$ 475e\textsuperscript{-}) compared to state of the art RD53B designs \cite{RD53BmeasGaioni} such as CROCv1\cite{Crocv1} and ITKv1\cite{ITKv1}. We also showed the implementation of machine learning-based approaches for data processing in the pixelated region, such as analyzing cluster shapes to determine track parameters and/or suppressing noise hits. By leveraging the proposed architecture, only useful clusters will be transmitted to the periphery for further processing and off-chip data transfer, thus minimizing the necessary data bandwidth for data collection under much higher collision rates for future upgrades. The second prototype chip is expected to be received in May 2024 and test results will be presented at the conference.

\newpage
\bibliographystyle{IEEEtran}
\bibliography{IEEEabrv,biblio}
\end{document}